\documentclass[12pt]{iopart}
\RequirePackage{amsbsy}
\RequirePackage{amssymb} 
\RequirePackage{mathrsfs}
\RequirePackage{epsfig}

\def\nn{\nonumber}
\def\a'{a^{\prime}}
\def\f'{f^{\prime}}
\def\h'{h^{\prime}}
\begin{document}
\title[Spontaneously broken conformal gravity]{Spontaneous breaking of conformal invariance in theories of conformally coupled matter and Weyl gravity}
\author{A. Edery$^1$, Luca Fabbri$^2$ and M. B. Paranjape$^3$}

\address{$^1$
Dept. of Physics, Bishop's University,
Lennoxville, Qu\'ebec, CANADA J1M 1Z7 }
\ead{aedery@ubishops.ca}
\address{$^2$Theory Group, INFN Section of Bologna,
Department of physics - University of Bologna,
Via Irnerio 46, C.A.P. 40126, Bologna, ITALY }
\ead{lucafabbri@bo.infn.it}
\address{$^3$ Groupe de physique des particules, Universit\'e de Montr\'eal,
C.P. 6128, succ. centre-ville, Montr\'eal, Qu\'ebec, CANADA  H3C 3J7}
\ead{paranj@lps.umontreal.ca}
\begin{abstract}
We study the theory of Weyl conformal gravity with matter degrees of freedom in a conformally invariant interaction. Specifically, we consider a triplet of scalar fields and $SO(3)$ non-abelian gauge fields, i.e. the Georgi-Glashow model conformally coupled to Weyl gravity. We show that the equations of motion admit solutions spontaneously breaking the conformal symmetry and the gauge symmetry, providing a mechanism for supplying a scale in the theory. The vacuum solution corresponds to anti-de-Sitter space-time, while localized soliton solutions correspond to magnetic monopoles in asymptotically anti-de-Sitter space-time. The resulting effective action gives rise to Einstein gravity and the residual $U(1)$ gauge theory. This mechanism strengthens the reasons for considering conformally invariant matter-gravity theory, which has shown promising indications concerning the problem of missing matter in galactic rotation curves.
\end{abstract}
\maketitle
\parskip=.5cm
\newpage
\section{Introduction}
Conformal gravity has been suggested as an alternative gravity theory which gives a partial resolution of the problem of missing matter in galaxies \cite{km} and \cite{m}. It is a simpler theory of gravity as the choice for the action, that is minimal in the number of derivatives, is unique and it has a higher degree of symmetry, invariance under local conformal transformations in addition to general covariance.  However, since it is a higher, fourth order theory of gravity, the integration of the equations of motion gives rise to two additional integration constants in comparison to the usual second order Einstein theory. Indeed, the solution analogous to the Schwarzschild solution corresponds to, in the weak field limit, the usual Newtonian potential augmented with a linear potential and a cosmological solution\cite{m}. The coefficient of the linear potential can be adjusted to become relevant at galactic scales. It is intriguing that a simpler theory of gravitation can gi!
 ve rise to a reasonable, alternative explanation for the question of missing matter. 

There is another avenue of interest for the conformally invariant theory of gravitation. Evidently, conformal invariance implies that there is no prescribed scale in the field theory, it has dimensionless coupling constant. As it is fourth order in derivatives, the propagator behaves as $\frac{1}{k^{4}}$ which helps with the perturbative ultraviolet renormalisability \cite{ste}. Indeed, it has been recently shown that conformal gravity is renormalisable to the 1 loop\cite{b-ps}.  It has been observed that theories with more symmetries are usually less divergent and more tractable as quantum field theories, there is some hope that conformal gravity might provide for a consistent theory of quantum gravity.

In this paper we study a particular theory of conformally invariant matter and gravitation. Even though it is a specific model we believe the results to be quite generic within this context. We find that the theory contains spontaneous breaking of the conformal invariance.  Conformally coupled matter/gravity and also spontaneous breaking of conformal invariance has been considered by Demir, Odintsov and Mannheim \cite{doem}.  Mannheim's work is closest in spirit to our paper.  He considers the conformally coupled scalar/gravitational field equations.  He finds for a negative scalar field self-coupling, a mass scale is generated in the theory and the metric corresponds to de-Sitter space-time.  We find that anti-de-Sitter space-time arises naturally as the vacuum solution of the theory for a positive scalar self coupling. This is interesting since AdS space-times have risen in importance over the last few years in the context of the correspondance between string theory and co!
 nformal field theory \cite{agmoo}. The value of the curvature of the AdS space-time is a constant but not fixed. Outside input must enter to stabilize the scale of the breaking of conformal invariance. We speculate this might come from the renormalization scale of the quantum corrections of the matter fields.

We also show numerical evidence for localized soliton solutions of the magnetic monopole type in the specific theory of the conformally coupled Georgi-Glashow \cite{gg} model. Similar solutions have been found for the case of Einstein gravity or string inspired gravity with Yang-Mills and Higgs fields, \cite{hbfrghs}, however no one has studied the case of conformal gravity. Our solutions have a non-singular core developing into an asymptotically anti-de-Sitter space-time. Such solutions would correspond to massive particles in the theory. 
\section{Conformally Invariant Action}
\subsection{Conformal Theory of Gravitation}
Conformal Gravity is a theory of gravitation, developed on the assumption of conformal symmetry and general covariance.  The geometrical structure used to translate the principle of covariance into a mathematical language is the tensorial calculus on pseudo-Riemannian differential manifolds, endowed with a metric and symmetric (i.e. torsionless) connection, in which the metric tensor is with negative signature: in the following, we will use the convention $(1,-1,-1,-1)$; for this geometrical theory, all local properties of the space are described by the Riemann tensor.
Riemann tensor is defined by the connection of the space; and the connection is given by the definition of the covariant derivative. 
In the following, we will represent the covariant derivative with respect to the coordinate $x^{\mu}$ with $\nabla_{\mu}$. Riemann curvature tensor will be represented by $R^{\beta}_{\phantom{\beta} \mu \nu \rho}$, one contraction gives the Ricci curvature tensor $R_{\mu \nu}=R^{\beta}_{\phantom{\beta} \mu \beta \nu}$ and further contraction gives the curvature scalar $R=R_{\mu \nu}g^{\mu \nu}$.  The tracefree part of the Riemann tensor, called the Weyl tensor $C^{\beta}_{\phantom{\beta} \mu \nu \rho}$, is shown to be conformally invariant (see, for example, \cite{wald}), that is, given any function $\Omega=\Omega(x)$, we can perform a transformation of the metric tensor as
\begin{eqnarray}
\nn
g'_{\mu \nu}=\Omega^{2}g_{\mu \nu}
\end{eqnarray}
for which indeed the Weyl tensor is invariant 
\begin{eqnarray}
\nn
C'^{\beta}_{\phantom{\beta} \mu \nu \rho}=C^{\beta}_{\phantom{\beta} \mu \nu \rho} \ .
\end{eqnarray}
Then, the tensorial calculus on pseudo-Riemannian differentiable manifolds in which all local properties of the space are described by Weyl tensor is what mathematically represents the principles of conformal invariance and general covariance. 

On the basis of such a theory, the action must be invariant under scale and coordinates trasformations, that is, it has to be a scalar built up using only Weyl tensor. 

Since Weyl tensor is tracefree, we cannot form any scalar by taking its contractions; the only way we have to obtain a scalar is to contract the products of two Weyl tensors: considering two Weyl tensors and contracting each index of one with an index of the other, we get only one possible choice for the conformally invariant scalar we are looking for.  Thus, the action defined as
\begin{eqnarray}
\nn
S=\int\sqrt{|g|}C_{\beta \mu \nu \rho}C^{\beta \mu \nu \rho}d^{4}x
\equiv \int\sqrt{|g|}C^{2}d^{4}x
\end{eqnarray}
is the unique action constructed using only the metric, which is invariant under (local) conformal and coordinate transformations; it will be our choice for the action of the gravitational field.

The action for a gravitational field in presence of other fields is obtained as follow
\begin{equation}
S=\int\sqrt{|g|}(C^{2}+kT)d^{4}x
\label{ac}
\end{equation}
where $k$ is a constant, with $T$ a scalar such that 
\begin{equation}
T'=\Omega^{-4}T\label{ct}
\label{transf}
\end{equation}
under conformal transformations.  When an explicit choice for the matter content is made, we must specify the conformal transformation properties of the corresponding fields to ensure the overall transformation property (\ref{ct}).  The tensor $W_{\mu \nu}$ is defined by
\begin{equation}
\delta(\sqrt{|g|}C^{2})\equiv-2\sqrt{|g|}W_{\mu \nu}\delta g^{\mu \nu}.
\end{equation}
\begin{eqnarray}
\nn
W_{\mu \nu}&=&\left(\frac{1}{6}g_{\mu \nu}\nabla^{2}R
+\frac{1}{3}\nabla_{\mu}\nabla_{\nu}R-\nabla^{2}R_{\mu \nu}\right) \\
&+&\frac{1}{3}R\left(2R_{\mu \nu}-\frac{1}{2}g_{\mu \nu}R\right)
+R^{\beta \rho}\left(\frac{1}{2}g_{\mu \nu}R_{\beta \rho}
-2R_{\beta \mu \rho \nu}\right)
\end{eqnarray}
is symmetric, tracefree and its covariant $4$-divergence vanishes.  The energy-momentum tensor is defined by  varying the term $T$ in the action with respect to the metric tensor
\begin{equation}
\delta(\sqrt{|g|}T) \equiv \frac{\sqrt{|g|}}{2}T_{\mu \nu}\delta g^{\mu \nu}.
\label{emt}
\end{equation}
$T_{\mu \nu}$ is symmetric, its $4$-divergence vanishes and, in the case of conformal theories, it is tracefree.  The action given in equation (\ref{ac}) yields the field equations 
\begin{equation}
W_{\mu \nu}=\frac{k}{4}T_{\mu \nu},
\end{equation}
evidently both sides are symmetric, tracefree and covariantly conserved.
\subsection{Gauge and Higgs fields in a Conformal Theory of Gravitation}
We consider a gauge potential as $n$ $4$-vectors $A_{\mu}^{(a)}$ (we put gauge group indices in parenthesis to avoid confusion with spatial indices).  This gives the field strength 
\begin{equation}
F_{\mu \nu}^{(a)}=\nabla_{\mu}A_{\nu}^{(a)}-\nabla_{\nu}A_{\mu}^{(a)}
+C^{a}_{\phantom{a}bc}A_{\mu}^{(b)}A_{\nu}^{(c)}
\end{equation}
where $C^{a}_{\phantom{a}bc}$ are the structure constants of the Lie algebra of the gauge group.  The field strength  transforms according to the adjoint representation of the gauge group. We introduce the Higgs boson, also in the adjoint representation, as $n$ scalars $\phi^{(a)}$. The gauge covariant derivative of Higgs bosons is given by
\begin{equation}
D_{\mu}\phi^{(a)}=\nabla_{\mu}\phi^{(a)}+C^{a}_{\phantom{a}bc}A_{\mu}^{(b)}\phi^{(c)}.
\end{equation}
The action for the Higgs bosons in interaction with the gauge field is given by the standard form
\begin{equation}
S=\int\sqrt{|g|}\left(\sum_{k=1}^{k=n}
D_{\mu}\phi^{(k)}D^{\mu}\phi^{(k)}-\frac{1}{4e^{2}}F_{\mu \nu}^{(k)}F^{\mu \nu (k)}\right)d^{4}x.
\label{act}
\end{equation}
We can extend the conformal transformations for the gauge fields as follow
\begin{equation}
A_{\mu}^{'(a)}=A_{\mu}^{(a)}
\end{equation}
which implies immediately
\begin{equation}
F_{\mu \nu}^{'(a)}=F_{\mu \nu}^{(a)}.
\end{equation}
and for the Higgs as 
\begin{equation}
\phi'^{(a)}=\Omega^{-1}\phi^{(a)}\label{h}
\end{equation}
which gives
\begin{equation}
(D_{\mu}\phi^{(a)})'=\Omega^{-1}(D_{\mu}\phi^{(a)}-\phi^{(a)}\nabla_{\mu}\ln{\Omega})\label{dh}.
\end{equation}
This implies that the action given in equation (\ref{act}) is not conformally invariant.
However, we can actually taking into account that the trace of  the curvature scalar is not conformally invariant  it is possible to couple it with Higgs fields, to give a conformally invariant action.  Such a modification was first considered in the context of massless free scalar fields by Penrose and G\"ursey \cite{pg}. The modifcation was further elaborated upon by Callan, Coleman and Jackiw \cite{ccj} with respect to the definition of the energy momentum tensor, and it has recently been further discussed by Jackiw \cite{j}.  For a general overview of conformal transformations see, for example, \cite{wald}. The transformation property of the  Higgs bosons given by equations (\ref{h},\ref{dh}) implies 
\begin{eqnarray}
\nn
(\sum_{a=1}^{a=n}D^{\mu}\phi^{(a)}D_{\mu}\phi^{(a)})'&
=&\Omega^{-4}\sum_{a=1}^{a=n}\left[\right.D^{\mu}\phi^{(a)}D_{\mu}\phi^{(a)}\\
\nn
&+&\phi^{(a)}(\phi^{(a)}\nabla_{\mu}\ln{\Omega}-
2\nabla_{\mu}\phi^{(a)})\nabla^{\mu}\ln{\Omega}\left.\right].
\end{eqnarray}
However the scalar curvature transforms as
\begin{eqnarray}
\nn
R'=\Omega^{-2}\left[(R
-(d-1)((d-2)\nabla_{\mu}\ln{\Omega}\nabla^{\mu}\ln{\Omega}
+2\nabla^{\mu}\nabla_{\mu}\ln{\Omega})\right].
\end{eqnarray}
Thus we can choose a term of the form
\begin{eqnarray}
\nn
\sum_{k=1}^{k=n}(\chi R\phi^{(k)}\phi^{(k)})
\end{eqnarray}
for which it is actually possible to get the conformal invariance. In $4$-dimensions, taking the factor $\chi$ equal to $1/6$ makes the sum 
\begin{eqnarray}
\nn
\sum_{k=1}^{k=n}(D_{\mu}\phi^{(k)}D^{\mu}\phi^{(k)}+\frac{1}{6}R\phi^{(k)}\phi^{(k)})
\end{eqnarray}
and consequently, the whole action, conformally invariant.
In addition, we can add the Higgs quartic self-coupling given as  $\lambda^{2}(\sum_{k=1}^{k=n}\phi^{(k)}\phi^{(k)})^{2}$, which is conformally invariant and a gauge invariant scalar by itself.

Finally, the conformally invariant action is given by
\begin{eqnarray}
\nn
S&=&\int\sqrt{|g|}(C^{2}+\sum_{k=1}^{k=n}(-\frac{1}{4e^{2}}F_{\mu \nu}^{(k)}F^{\mu \nu (k)}\\
&+&D_{\mu}\phi^{(k)}D^{\mu}\phi^{(k)}+\phi^{(k)}\phi^{(k)}(\frac{1}{6}R-\lambda^{2}\sum_{j=1}^{j=n}\phi^{(j)}\phi^{(j)})))d^{4}x
\label{acti}
\end{eqnarray}
where $e$ is the gauge coupling constant; this action describes gauge fields in interaction with Higgs bosons in conformally invariant gravitation.
\subsection{Spontaneous Conformal Symmetry Breaking and Anti-de-Sitter Spaces}
We can introduce the compact notation for which $\sum_{k=1}^{k=n}\phi^{(k)}\phi^{(k)}=\phi^{2}$; we see that in the action (\ref{acti}) the last two terms do not contain derivatives of the scalar field, therefore, they can be seen as a non-minimal potential for the scalar field
\begin{equation}
V(\phi)=-\frac{1}{6}R\phi^{2}+\lambda^{2}\phi^{4}.
\end{equation}
This is not the potential usually considered in Higgs mechanism, since the term $\phi^{2}$ is proportional to the scalar curvature $R$, which is not the mass of Higgs field, nor a constant; this non-minimal potential can provide for spontaneous breaking of the conformal and gauge symmetry.  Non-constant curvatures produce unusual non-local structures as has been recently found in \cite{gs}).  We confine ourselves, in this paper,  to the classical aspects of the theory.  We consider the case of positive quartic scalar self-coupling, $\lambda^2>0$.  This case is required for stability in flat backgrounds.

The case for which the scalar curvature is a constant corresponds to maximally symmetric spaces (see, for example, \cite{ma}). The constant scalar curvature is then given by
\begin{eqnarray}
\nn
R=bd(d-1)
\end{eqnarray}
where $d$ is the dimension and $b$ a constant: when $b=0$ the space-time is simply Minkowski space, when $b<0$ the space-time is called de-Sitter space and for $b>0$ the space-time is called anti-de-Sitter (AdS) space.  (Note that these identifications are convention dependent.)
For $b<0$ the only a stable equilibrium is with $\phi^{2}=0$; on the other hand for $b>0$ a stable equilibrium is obtained with $\phi^{2}\neq 0$, i.e. symmetry breaking corresponds to positive values of the constant scalar curvature, namely, in AdS space-times.

We will make the \emph{Ansatz} that the AdS geometry is only asymptotical, that is only the boundary of the space-time is geometrically an AdS space. 
In asymptotically AdS space-times, then, we will set the numerical value of the scalar curvature at the boundary of the space-time to be equal to $12\lambda^{2}v^{2}$, with $v$ an arbitrary constant.  In this case, $\phi^{2}=v^{2}$ is the solution which asymptotically gives the spontaneous breakdown of the gauge symmetry, giving mass to some of the gauge fields via the Higgs mechanism. It is important to stress the fact that gauge symmetry breaking produces massive fields, which are no longer conformally invariant, thus it also implies that the conformal symmetry is broken in the process.  

The non minimal potential for the scalar field also serves to induce the Einstein action for the metric.  Indeed, if the scalar field is non zero in a local region, we can choose the  conformal gauge
\begin{equation}
\phi(x)\rightarrow \phi_0
\end{equation}
where $\phi_0$ is an arbitrary, non zero constant.  Then the net effect of the non minimal scalar potential is to induce the Einstein action for the metric fluctuations, in a conformal to  de Sitter background.
\section{$SO(3)$ Symmetries}
\subsection{Georgi-Glashow model and $SO(3)$ gauge symmetry}
In the Georgi-Glashow model the basic assumption is that we have a triplet of Higgs bosons and a triplet of gauge fields for which the symmetry group is $SO(3)$, namely, the structure constants of the Lie algebra are the completely anti-symmetric $3$-dimensional coefficients of Ricci-Levi-Civita 
\begin{equation}
C^{abc}=\varepsilon^{abc}.
\end{equation}
The field strength is given by
\begin{equation}
\nn
F_{\mu \nu}^{(a)}=\nabla_{\mu}A_{\nu}^{(a)}-\nabla_{\nu}A_{\mu}^{(a)}
+\varepsilon^{a}_{\phantom{a}bc}A_{\mu}^{(b)}A_{\nu}^{(c)}
\end{equation}
and the gauge covariant derivative for Higgs bosons is given by
\begin{equation}
\nn
D_{\mu}\phi^{(a)}=\nabla_{\mu}\phi^{(a)}+\varepsilon^{a}_{\phantom{a}bc}A_{\mu}^{(b)}\phi^{(c)}.
\end{equation}

We can introduce the $3$-dimensional notation, for which we can put
\begin{equation}
\nn
\sum_{k=1}^{k=3}V^{(k)}W^{(k)}=\vec{V} \cdot \vec{W}
\end{equation}
and
\begin{equation}
\nn
(\vec{V} \wedge \vec{W})^{(a)}=\varepsilon^{a}_{\phantom{a}bc}V^{(b)}W^{(c)}.
\end{equation}

With this notation, we have 
\begin{equation}
\vec{F}_{\mu \nu}=\nabla_{\mu}\vec{A}_{\nu}-\nabla_{\nu}\vec{A}_{\mu}+\vec{A}_{\mu}\wedge\vec{A}_{\nu}
\end{equation}
and the gauge covariant derivative for Higgs bosons is given by
\begin{equation}
D_{\mu}\vec{\phi}=\nabla_{\mu}\vec{\phi}+\vec{A}_{\mu}\wedge\vec{\phi}.
\end{equation}
Then, we can set 
\begin{equation}
\sum_{k=1}^{k=n}\phi^{(k)}\phi^{(k)}\equiv \vec{\phi} \cdot \vec{\phi}=\phi^{2}
\end{equation}
and
\begin{equation}
\sum_{k=1}^{k=n}D_{\mu}\phi^{(k)}D^{\mu}\phi^{(k)}\equiv 
D_{\mu}\vec{\phi} \cdot D^{\mu}\vec{\phi}=(D\phi)^{2}
\end{equation}
and also
\begin{equation}
\sum_{k=1}^{k=n}F_{\mu \nu}^{(k)}F^{\mu \nu (k)} \equiv
\vec{F}^{\mu \nu} \cdot \vec{F}_{\mu \nu} = F^{2}.
\end{equation}
Finally, we have 
\begin{equation}
D_{\mu}\vec{F}^{\mu \nu}=\nabla_{\mu}\vec{F}^{\mu \nu}+\vec{A}_{\mu}\wedge\vec{F}^{\mu \nu}
\end{equation}
and
\begin{eqnarray}
\nn
D^{2}\vec{\phi}&=&D^{\mu}D_{\mu}\vec{\phi}=\nabla^{\mu}\nabla_{\mu}\vec{\phi}\\
&+&2\vec{A}^{\mu}\wedge\nabla_{\mu}\vec{\phi}+\vec{A}_{\mu}\wedge(\vec{A}^{\mu}\wedge\vec{\phi})+
\nabla_{\mu}\vec{A}^{\mu}\wedge\vec{\phi}.
\end{eqnarray}
Furthermore, we can define the N\"other's charges for the triplet of scalars as follow
\begin{equation}
\vec{J}_{\mu}=D_{\mu}\vec{\phi}\wedge\vec{\phi}
\end{equation}

The total energy-momentum tensor is defined according to (\ref{emt})
\begin{eqnarray}
\nn
T_{\mu \nu}&=&\frac{1}{e^{2}}\left(\right.\frac{1}{4}g_{\mu \nu}F^{2}
-\vec{F}_{\mu \beta}\cdot\vec{F}_{\nu}^{\phantom{\nu} \beta}\left.\right)\\
\nn
&+&2D_{\mu}\vec{\phi} \cdot D_{\nu}\vec{\phi}-
g_{\mu \nu}((D\phi)^{2}-\lambda^{2}\phi^{4})\\
&+&\frac{1}{3}\left((g_{\mu \nu}\nabla^{2}\phi^{2}-\nabla_{\mu}\nabla_{\nu}\phi^{2})+
(R_{\mu \nu}-\frac{1}{2}g_{\mu \nu}R)\phi^{2}\right)
\end{eqnarray}
However, we can isolate the contribution due to the gauge field alone
\begin{equation}
T^{(gauge)}_{\mu \nu}=\frac{1}{e^{2}}\left(\right.\frac{1}{4}g_{\mu \nu}F^{2}
-\vec{F}_{\mu \beta}\cdot\vec{F}_{\nu}^{\phantom{\nu} \beta}\left.\right)
\end{equation}
and the remaining part is then written as
\begin{eqnarray}
\nn
\Theta_{\mu \nu}&=&2D_{\mu}\vec{\phi} \cdot D_{\nu}\vec{\phi}-
g_{\mu \nu}((D\phi)^{2}-\lambda^{2}\phi^{4})\\
\nn
&+&\frac{1}{3}\left((g_{\mu \nu}\nabla^{2}\phi^{2}-\nabla_{\mu}\nabla_{\nu}\phi^{2})+
(R_{\mu \nu}-\frac{1}{2}g_{\mu \nu}R)\phi^{2}\right)
\end{eqnarray}
We point out that now all the dependence on the scalar field is in the tensor $\Theta_{\mu\nu}$, which contains terms related to the interaction of the Higgs fields with the gauge field and gravity one (for further discussion, see, for example \cite{ccj} and \cite{j}). 
Given the gauge symmetry group, and the $3$-dimensional notation, it is possible to write the action 
\begin{equation}
S=\int\sqrt{|g|}(C^{2}-\frac{1}{4e^{2}}F^{2}+(D\phi)^{2}+\frac{1}{6}R\phi^{2}-\lambda^{2}\phi^{4})d^{4}x.
\end{equation}
\subsection{Stationary and spherically symmetric spaces}
The geometrical configuration of the spacetime we want to study is chosen to be stationary and spherically symmetric, i.e. the distribution of energy does not depend on time and it is isotropically distributed around the origin, hence, the symmetry group of the spatial isometries is $SO(3)$.

This structure for spatial isometries allows us to consider $3$ linearly independent $4$-dimensional Killing vectors in spherical coordinates $(t,r,\theta,\varphi)$, given as follows
\begin{eqnarray}
\nn
\xi_{(1)}=(0,0,\cos{\varphi},-\sin{\varphi}\cot{\theta})
\end{eqnarray}
\begin{eqnarray}
\nn
\xi_{(2)}=(0,0,-\sin{\varphi},-\cos{\varphi}\cot{\theta})
\end{eqnarray}
\begin{eqnarray}
\nn
\xi_{(3)}=(0,0,0,1)
\end{eqnarray}
for which
\begin{eqnarray}
\nn
[\xi_{(i)},\xi_{(j)}]=\varepsilon_{ijk}\xi_{(k)}
\end{eqnarray}
the correct representation of the symmetry group for $SO(3)$.

The line element in spherical coordinates has to be isotropic, that is there exists a system of coordinates in which it has the form
\begin{equation}
\nn
ds^{2}=A(r)dt^{2}-B(r)dr^{2}-r^{2}(d\theta^{2}+\sin^{2}{\theta}d\varphi^{2})
\end{equation}
It has been noted by Mannheim and Kazanas (\cite{km}) that, via a sequence of coordinate and conformal transformations, one can always bring the general static, spherically symmetric metric to a form in which $A=1/B$.  Thus it is justified to set $A=1/B=1+h(r)$ for a smooth function $h(r)$, so that
\begin{equation}
ds^{2}=(1+h(r))dt^{2}-\left(\frac{1}{1+h(r)}\right)dr^{2}-r^{2}(d\theta^{2}+\sin^{2}{\theta}d\varphi^{2}).
\end{equation}
Given this line element, we can compute the curvature tensors; the Riemann tensor has all the components equal to zero except for
\begin{equation}
\nn
R_{trtr}=-\frac{h''}{2}
\end{equation}
\begin{equation}
\nn
R_{t\theta t\theta}=-\frac{r(1+h)h'}{2}
\end{equation}
\begin{equation}
\nn
R_{r\theta r\theta}=\frac{rh'}{2(1+h)}
\end{equation}
\begin{equation}
\nn
R_{t\varphi t\varphi}=-\frac{r(1+h)h'}{2}\sin^{2}{\theta}
\end{equation}
\begin{equation}
\nn
R_{r\varphi r\varphi}=\frac{rh'}{2(1+h)}\sin^{2}{\theta}
\end{equation}
\begin{equation}
\nn
R_{\varphi \theta \varphi \theta}=r^{2}h\sin^{2}{\theta},
\end{equation}
while the Ricci tensor has all the component vanishing except for
\begin{equation}
\nn
R_{tt}=(1+h)(\frac{h''}{2}+\frac{h'}{r})
\end{equation}
\begin{equation}
\nn
R_{rr}=-\frac{1}{(1+h)}(\frac{h''}{2}+\frac{h'}{r})
\end{equation}
\begin{equation}
\nn
R_{\theta \theta}=-(rh'+h)
\end{equation}
\begin{equation}
\nn
R_{\varphi \varphi}=-(rh'+h)\sin^{2}{\theta}
\end{equation}
and finally, the scalar curvature is given by
\begin{equation}
\nn
R=\frac{(r^{2}h)''}{r^{2}},
\end{equation}
and in these coordinates, the asymptotic condition on the scalar field is
\begin{equation}
\lim_{r \to \infty}\left(2\frac{h}{r^{2}}+4\frac{h'}{r}+h''\right)=12\lambda^{2}v^{2}
\end{equation}
\subsection{Spatial and gauge mixed $SO(3)$ symmetry}
In the following, we will make an \emph{Ansatz} based on the idea that the lowest energy solutions are those with the maximal symmetry (see \cite{t} and \cite{p} and also, for example, \cite{cl}), which is $SO(3)$.  We note that the spatial symmetry $SO(3)$ is actually the same we as the gauge symmetry.
The explicit assumptions on the components of the fields are (in cartesian coordinates), for the gauge fields: 
\begin{equation}
A^{0 (a)}=0
\end{equation}
\begin{equation}
A^{i (a)}=q(r^{2})\varepsilon^{aik}r^{k}
\end{equation}
for the Higgs field:
\begin{equation}
\phi^{(a)}=f(r^{2})\frac{r^{a}}{r}
\end{equation}
where $f(r)$ and $q(r)$ have to be fixed by the field equations. In terms of the $3$-dimensional notation they are
\begin{equation}
\vec{A}_{0}=0
\end{equation}
\begin{equation}
\vec{A}_{a}=q(r^{2})\vec{n}_{a}\wedge\vec{r}
\end{equation}
where $\vec{n}_{a}$ is the unity vector along the $a$ axis while for the Higgs 
\begin{equation}
\vec{\phi}=f(r^{2})\frac{\vec{r}}{r}.
\end{equation}
Since the frame we have chosen is in spherical coordinates, we must write all of the fields in spherical coordinates $(r,\theta,\varphi)$. 
For the scalar fields a simple calculation gives
\begin{equation}
\nn
\phi^{(1)}=f(r^{2})\sin{\theta}\sin{\varphi}
\end{equation}
\begin{equation}
\nn
\phi^{(2)}=f(r^{2})\sin{\theta}\cos{\varphi}
\end{equation}
\begin{equation}
\nn
\phi^{(3)}=f(r^{2})\cos{\theta}.
\end{equation}
For the gauge fields we notice that the explicit expression for the Killing vectors for $SO(3)$ in Cartesian coordinates is given by
\begin{eqnarray}
\nn
\xi_{(1)}=
\left(\begin{array}{c}
0 \\
0 \\
z \\ 
-y
\end{array}\right); \ \
\xi_{(2)}=
\left(\begin{array}{c}
0 \\
-z \\
0 \\ 
x
\end{array}\right); \ \
\xi_{(3)}=
\left(\begin{array}{c}
0 \\
y \\
-x \\ 
0
\end{array}\right)
\end{eqnarray}
and these are proportional to the explicit expressions of the gauge fields themselves; thus we find the conditions
\begin{eqnarray}
\nn
q\xi_{(i)}^{\mu}=A_{(i)}^{\mu}.
\end{eqnarray}
These conditions immediately give us all the properties of $\vec{A}$, knowing $\vec{\xi}$: thus the expression for the gauge potentials in spherical coordinates is easily seen as
\begin{equation}
\nn
A_{(1)}=
q(r^{2})\left(\begin{array}{c}
0 \\
0 \\
\cos{\varphi} \\ 
-\sin{\varphi}\cot{\theta}
\end{array}\right)
\end{equation}
\begin{equation}
\nn
A_{(2)}=
q(r^{2})\left(\begin{array}{c}
0 \\
0 \\
-\sin{\varphi} \\ 
-\cos{\varphi}\cot{\theta}
\end{array}\right)
\end{equation}
\begin{equation}
\nn
A_{(3)}=
q(r^{2})\left(\begin{array}{c}
0 \\
0 \\
0 \\ 
1
\end{array}\right).
\end{equation}
Now we are able to write down explicitly the fields strength and the gauge covariant derivatives of this theory; the most important are listed as follow, defining
\begin{equation}
1+r^{2}q(r^{2})=a(r^{2}),
\end{equation}
we get for the field strength 
\begin{equation}
\vec{F}_{t r}=0
\end{equation}
\begin{equation}
\vec{F}_{t \theta}=0
\end{equation}
\begin{equation}
\vec{F}_{t \varphi}=0
\end{equation}
\begin{equation}
\vec{F}_{r \theta}=\frac{a'}{a-1}\vec{A}_{\theta}
\end{equation}
\begin{equation}
\vec{F}_{r \varphi}=\frac{a'}{a-1}\vec{A}_{\varphi}
\end{equation}
\begin{equation}
\vec{F}_{\theta \varphi}=\frac{1-a^{2}}{f}\sin{\theta}\vec{\phi}.
\end{equation}
The non-abelian electric and magnetic fields are given by
\begin{equation}
\vec{E}_{a}=0
\end{equation}
and
\begin{equation}
\vec{B}^{\theta}=-\frac{1}{r^{2}\sin{\theta}}\frac{a'}{(a-1)}\vec{A}_{\varphi}
\end{equation}
\begin{equation}
\vec{B}^{\varphi}=\frac{1}{r^{2}\sin{\theta}}\frac{a'}{(a-1)}\vec{A}_{\theta}
\end{equation}
\begin{equation}
\vec{B}^{r}=\frac{1-a^{2}}{r^{2}f}\vec{\phi}
\end{equation}
in which the arrow represents the vectors in the internal space as always. The non-abelian magnetic field has a radial part that is parallel in internal space to the scalar field.

With this, we can write the explicit form for the terms in the action
\begin{equation}
D_{\mu}\vec{\phi} \cdot D^{\mu}\vec{\phi}=-((1+h)(f')^{2}+2\frac{a^{2}f^{2}}{r^{2}})
\end{equation}
and
\begin{equation}
\vec{F}_{\alpha \beta}\cdot \vec{F}^{\alpha \beta}
=\frac{2}{r^{2}}(2(1+h)(a')^{2}+\frac{(a^{2}-1)^{2}}{r^{2}})
\end{equation}
so, we have the complete expression in spherical coordinates for both the gauge and the Higgs fields, giving the whole action in spherical coordinates:
\begin{eqnarray}
\nn
S&=&-4\pi \int_{0}^{+\infty}\mathscr{L}dr=-4\pi \int_{0}^{+\infty}\left[\right.(\frac{a'^{2}}{e^{2}}+\frac{(a^{2}-1)^{2}}{2r^{2}e^{2}})\\
\nn
&+&(\frac{ha'^{2}}{e^{2}})+(r^{2}f'^{2})+(2a^{2}f^{2})+(hr^{2}f'^{2}+\lambda^{2}r^{2}f^{4}-f^{2}\frac{(r^{2}h)''}{6}\\
&+&\frac{1}{3}(-r^{2}h''^{2}+8h'^{2}-4\frac{h^{2}}{r^{2}}+4rh'h''+8\frac{hh'}{r}+8hh'')
\left.\right]dr .\label{fe}
\end{eqnarray}
\section{Field equations}
We are now able to obtain the field equations of the theory, (varying the action with respect to the scalar fields, the vector fields and the metric; respectively,).  For Higgs field we get
\begin{equation}
D^{2}\vec{\phi}=\left(\frac{R}{6}-2\lambda^{2}\phi^{2}\right)\vec{\phi}
\end{equation}
for gauge fields
\begin{equation}
D_{\mu}\vec{F}^{\mu \nu}=2e\vec{J}^{\nu}
\end{equation}
and for the metric
\begin{equation}
W_{\mu \nu}=\frac{1}{4}T_{\mu \nu}.
\end{equation}
The coupling constant $k$ is absorbed into the coupling constants $e$ and $\lambda^2$ and into the scalar field by an appropriate rescaling.
One can confirm the tracelessnes of the energy-momentum tensor 
\begin{equation}
\nn
T^{\mu}_{\phantom{\mu} \mu}=0
\end{equation}
as it should be in a conformal theory and also see that the conservation laws are satisfied
\begin{equation}
\nn
\nabla_{\mu}T^{\mu \nu}=0
\end{equation}
and
\begin{equation}
\nn
D_{\mu}\vec{J}^{\mu}=0 .
\end{equation}
The whole system of field equations then is
\begin{equation}
D^{2}\vec{\phi}=\left(\frac{R}{6}-2\lambda^{2}\phi^{2}\right)\vec{\phi}
\end{equation}
\begin{equation}
D_{\mu}\vec{F}^{\mu \nu}=2e\vec{J}^{\nu}
\end{equation}
\begin{equation}
W_{\mu \nu}=\frac{1}{4}T_{\mu \nu}
\end{equation}
and, in the last equation, spherical coordinates and spherical symmetry imply that only the diagonal equations are non-trivial. Furthermore, we have $W^{\varphi}_{\phantom{\varphi} \varphi}=W^{\theta}_{\phantom{\theta} \theta}$ and $T^{\varphi}_{\phantom{\varphi} \varphi}=T^{\theta}_{\phantom{\theta} \theta}$, hence, only one of the angular equations is independent. Using the constraint of tracelessness we can, in fact, remove this angular equation, leaving only the temporal and the radial equations. Finally it can be proven (see \cite{km} and \cite{m}) that only one of the these two remaining equations is independent: so, we can choose a linear combination of the radial and the temporal one, to get the field equation for gravity
\begin{equation}
(rh)''''=\frac{r}{2}(2f'^{2}-ff'')+\frac{3a'^{2}}{2re^{2}}.
\end{equation}
Here the prime is the derivative with respect to  $r$.
The symmetries imposed reduce the Higgs field equations to the one independent equation
\begin{equation}
((1+h)r^{2}f')'=f(2a^{2}+r^{2}(2\lambda^{2}f^{2}-\frac{(r^{2}h)''}{6r^{2}}))
\end{equation}
and the gauge field equations reduce to 
\begin{equation}
((1+h)a')'=a(2f^{2}e^{2}+\frac{a^{2}-1}{r^{2}}).
\end{equation}
Thus, the entire set of field equations is
\begin{equation}
(rh)''''=\frac{r}{2}(2f'^{2}-ff'')+\frac{3a'^{2}}{2re^{2}}\label{e1}
\end{equation}
\begin{equation}
((1+h)r^{2}f')'=f(2a^{2}+r^{2}(2\lambda^{2}f^{2}-\frac{(r^{2}h)''}{6r^{2}}))\label{e2}
\end{equation}
\begin{equation}
((1+h)a')'=a(2f^{2}e^{2}+\frac{a^{2}-1}{r^{2}})\label{e3}
\end{equation}
containing the three unknown functions $h$, $f$, $a$. This result can also be obtained by varying the $SO(3)$ symmetric action  (\ref{fe}) with respect to  $h$, $f$ and $a$. 
\section{Solutions}
\subsection{External solutions}
We define the core of the monopole as that region in which we have nonzero density of N\"other's charge. The external solution is considered to be the solution outside this definition of the core.  Outside the core the vanishing  N\"other's density allows a non-trivial solution $a=0$. This is non-trivial because there is a long range monopole like magnetic field.  It is possible to solve the field equations exactly for this external solution.   A solution which supportts spontaneous symmetry breaking of the conformal symmetry simply requires that $f$ is not identically $0$.  We are interested in solutions of the form 
$f=v$. Given this, the only possible solution for the metric is given as 
$h=\frac{k}{r}+\lambda^{2}v^{2}r^{2}$, which recovers a part of the general vacuum solution found by Mannheim and Kazanas (e.g. \cite{km}).

This solution can be summarized as follow
\begin{equation}
a(r)=0
\label{sol1}
\end{equation}
\begin{equation}
f(r)=v
\label{sol2}
\end{equation}
\begin{equation}
h(r)=\frac{-\beta}{r}+\lambda^{2}v^{2}r^{2}
\label{sol3}
\end{equation}
for any value of the free parameter $\beta$ and for any positive value of the parameter $v$, which can be changed by conformal scaling.

In addition, note that the full form of the Mannheim-Kazanas matter free solution, $h(r)= \alpha -\beta/r +\gamma r +\delta r^2$, can be recovered by taking $a(r)=0$ but allowing $f(r)=1/(pr+q)$ for $p,q$ constants.  This makes the term $(2f'^{2}-ff'')$ in equation (\ref{e1}) vanish. When $q=0$  the coefficients $\alpha, \beta, \gamma, \delta$ are are given by  $\alpha=6\lambda^{2}/p^{2}$, $\beta=0$, while $\gamma , \delta$ are unconstrained.  When $q\neq0$ we get $\alpha = -3\beta p/q$, $\gamma =2p/q-3\beta p^{2}/q^{2}$ and $\delta= \lambda^2/q^2 +p^2/q^2 -\beta p^3/q^3$ with $\beta$ a free parameter.  

If instead we assume that the scalar field takes the value which minimizes the scalar potential $2\lambda^{2}f^{2}-\frac{(r^{2}h)''}{6r^{2}}=0$, which is properly seen as a stabilty criterion, the solutions are given by,  for  $q=0$,  $h(r)= 6 \lambda^2/p^2$ and $f(r)=1/pr$, and our preferred solution, for $q\ne 0$, $f=1/q$, $p=0$ and $h=-\beta/r+\lambda^2r^2/q^{2}$.  When $q=1/v$ the solution given above in equations (\ref{sol1})-(\ref{sol3}) is reproduced.

We emphasize that all the previous considerations hold away from singularities, since the fourth order derivative of the function $h$ is the radial component of a fourth order Laplacian, an operator whose action on a $1/r$ term yields the derivative of a delta function (see e.g. \cite{maka}).

In the solution (\ref{sol3}) we can see the presence of the Schwarzschild solution, which allows us to recover Einstein  gravitation at short  distances.  The quadratic term,  which is dominant at large distances,  gives rise to a constant, positive scalar curvature for the positive $\lambda^{2}v^{2}$.  Hence, the space-time is asymptotically  anti-de-Sitter. We note that this solution does \emph{not} reproduce the part $\sigma^{2}/r^{2}$ of the Reissner-Nordstrom solution corresponding to charged, spherically symmetric distributions of matter. The reason is simple: in conformal gravitation the equation of motion for the metric components is fourth order in derivatives, not second order as in the Einstein theory.  The integration of these equations precludes the appearance of the Reissner-Nordstrom behaviour.  

The Higgs field solution (\ref{sol2}) is a vector in the $3$-dimensional internal space, with a (positive) constant norm and radial direction, therefore topologically non-trivial. 
Finally, the feature that the ``gauge field'' $a$ in the solution (\ref{sol1}) vanishes means that the original gauge field $q$ behaves as $1/r^{2}$. This, together with the topologically non-trivial Higgs field, leads us to the fact that the gauge fields give rise to a non-zero magnetic field strength that is purely radial, and proportional to the Higgs field in the internal space:
\begin{equation}
B^{(k)r}=\frac{\phi^{(k)}}{v}\frac{1}{r^{2}}
\end{equation}

\subsection{Plot of the solutions}
For the problem concerning a more general solution, i.e. a solution which can take into account also the internal behaviour as well as the external one, we can try to find a solution via numerical calculations. The latter field equations need eight initial conditions to be given at the origin which specify a unique solution. A simple Mathematica program finds the solutions of the type that we are looking for.

A set of initial conditions can be chosen to give the solution
\begin{figure}[h]
\begin{center}
\includegraphics[width=12cm]{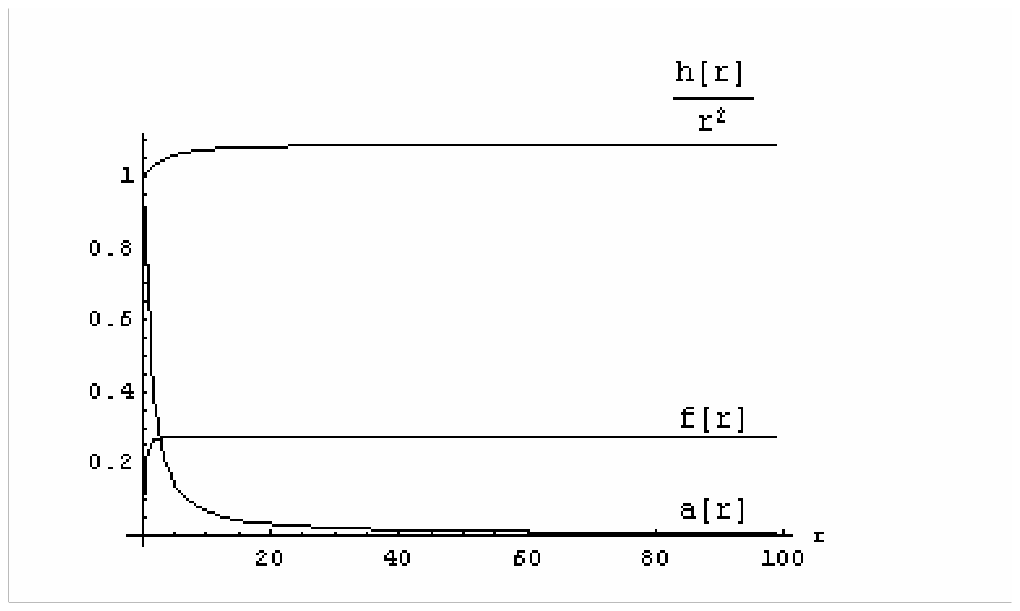}
\label{fig: Plot of the Solutions}
\caption{Plot of $a(r)$, $f(r)$ and $h(r)/r^2$ for $e=1$.}
\end{center}
\end{figure}

It is worthwhile to point out that the function $a$ related to the gauge field tends to vanish as the radius goes to infinity; correspondingly the scalar field tends to a constant value, determined by the asymptotical quadratic behaviour of the gravitational conformal field.
\section{Conclusions}
Conformal gravity is an interesting alternative theory of gravitation.  It is simpler than Einstein gravity since it does not have a scale, it has a unique action in minimal number of derivatives and it has greater symmetry.  However, the lack of conformal invariance in our observable universe must be addressed if conformal gravity is to be seriously considered as an alternative theory of gravitation.  The conformal invariance could be spontaneously broken, explicitly broken or anomalously broken as elucidated by Mannheim \cite{ma}.  In this paper, we have studied  a classical theory coupling matter and gravitational degrees of freedom in a conformally invariant manner.   Spontaneous symmetry breaking produces a scale in the theory.  The actual value of this scale is not fixed by the spontaneous symmetry breaking.   Mannheim showed that a nonzero VEV for the scalar field conformally coupled to Weyl gravity is consistent with a de-Sitter geometry for a negative  and anti-de-Sitter geometry for positive, quartic scalar self-coupling.  We have studied the situation where the VEV is consistent with anti-de-Sitter geometry with a positive scalar self-coupling.  We have additionally been able to find localized, topological soliton solutions in this theory.   These solitons represent  massive, particle states in the theory. This gives us some optimism and avenues for further research in the conformal theory of gravitation.
\ack 
We thank NSERC of Canada and INFN-Bologna for financial support. We also thank Silvio Bergia for helpful discussions.


\section*{References}
\begin{thebibliography}{30}
\bibitem{km}
P.~D.~Mannheim and D.~Kazanas, 
\textit{Astrophys. J.} \textbf{342}, 635 (1989).
\bibitem{m}
P.~D.~Mannheim,
\textit{Found. Phys.} \textbf{24}, 487 (1994).
\bibitem{ste}
K.~S.~Stelle,
\textit{Phys. Rev. D} \textbf{16}, 953 (1977).
\bibitem{b-ps}
G.~Berredo-Peixoto and I.~L.~Shapiro,
\textit{Phys. Rev. D} \textbf{70}, 044024 (2004).
\bibitem{doem}  
D.~Demir, 
\textit{Phys. Lett.} \textbf{B584}, 133 (2004);\\
S.~Odintsov, 
\textit{Phys. Lett.} \textbf{B336}, 347 (1994).
\bibitem{agmoo}  
O.~Aharony, S.~S.~Gubser, J.~M.~Maldacena, H.~Ooguri and Y.~Oz,  
\textit{Phys. Rept.} \textbf{323}, 183 (2000).
\bibitem{gg}
H.~Georgi and S.~L.~Glashow,
\textit{Phys. Rev. Lett.} \textbf{28}, 1494 (1972).
\bibitem{hbfrghs}
J.~Bjoraker and Y.~Hosotani,
\textit{ Phys. Rev. Lett.} \textbf{84}, 1853 (2000);\\ 
P.~Forgacs and S.~Reuillon, 
\textit{Phys. Rev. Lett.} \textbf{95}, 061101 (2005);\\ 
D.~Garfinkle, G.~T.~Horowitz and A.~Strominger, 
\textit{Phys. Rev. D} \textbf{43}, 3140 (1991)
[Erratum-ibid. \textbf{45}, 3888 (1992)];\\
G.~W.~Gibbons and K.~I.~Maeda,
\textit{Nucl. Phys.} \textbf{B298}, 741 (1988).
\bibitem{wald}
R.~M.~Wald,
``General Relativity'',
Chicago University Press, Chicago (1984).
\bibitem{pg} 
R.~Penrose, 
\textit{Proc. Roy. Soc. Lond.} \textbf{A284}, 159 (1965);\\
F.~G\"ursey, 
\textit{Ann. Phys.} \textbf{24}, 211 (1963).
\bibitem{ccj} 
C.~G.~Callan, S.~Coleman and R.~Jackiw, 
\textit{Ann. Phys.} \textbf{59}, 42 (1970).
\bibitem{j} 
R.~Jackiw, 
hepth/0511065, (2005).
\bibitem{gs} 
E.~V.~Gorbar and I.~L.~Shapiro,
JHEP \textbf{02}, 060 (2004).
\bibitem{ma} 
P.~D.~Mannheim, 
GRG \textbf{22}, 289 (1990).
\bibitem{t}
G.~'t Hooft
\textit{Nucl. Phys.} \textbf{B79}, 276 (1974).
\bibitem{p}
A.~M.~Polyakov
\textit{Sov. Phys.} JETP \textit{Lett.}\textbf{20}, 194 (1974).
\bibitem{cl}
Ta-Pei Cheng and Ling-Fong Li,
``Gauge theory of elementary particle physics'',
Oxford University Press, New York (1984).
\bibitem{maka}
P.~D.~Mannheim and D.~Kazanas, 
GRG \textbf{26}, 337 (1994).
\end{thebibliography}
\end{document}